\input harvmac
\input graphicx
\def\Title#1#2{\rightline{#1}\ifx\answ\bigans\nopagenumbers\pageno0\vskip1in
\else\pageno1\vskip.8in\fi \centerline{\titlefont #2}\vskip .5in}
%

%
%
%
%
\ifx\includegraphics\UnDeFiNeD\message{(NO graphicx.tex, FIGURES WILL BE IGNORED)}
\def\figin#1{\vskip2in}
\else\message{(FIGURES WILL BE INCLUDED)}\def\figin#1{#1}
\fi
\def\Fig#1{Fig.~\the\figno\xdef#1{Fig.~\the\figno}\global\advance\figno
 by1}
%
%
%
%
\def\Ifig#1#2#3#4{
\goodbreak\midinsert
\figin{\centerline{
\includegraphics[width=#4truein]{#3}}}
\narrower\narrower\noindent{\footnotefont
{\bf #1:}  #2\par}
\endinsert
}
%
%
\font\ticp=cmcsc10

\def\ajou#1&#2(#3){\ \sl#1\bf#2\rm(19#3)}
\def\jou#1&#2(#3){,\ \sl#1\bf#2\rm(19#3)}
\def\hf{{1\over 2}}

\def\frac#1#2{{#1\over#2}}

\def\eg{{\it e.g.,}}

\def\cL{{\cal L}}

\def\cR{{\cal R}}
\def\cC{{\cal C}}

\def\ie{{\it i.e.,}}
\def\la{\lambda}
%
%
\lref\PolchinskiSM{
J.~Polchinski and A.~Strominger,
``New Vacua for Type II String Theory,''
Phys.\ Lett.\ B {\bf 388}, 736 (1996)
[arXiv:hep-th/9510227].
}
\lref\BarrowIA{
J.~D.~Barrow,
``Cosmic No Hair Theorems And Inflation,''
Phys.\ Lett.\ B {\bf 187}, 12 (1987).
}
\lref\BBL{J.D. Barrow, A.B. Burd, and D. Lancaster,
``Three-dimensional classical spacetimes," Class. Quant. Grav.
{\bf 3}, 551 (1986).
}
\lref\Halliwell{J.~J.~Halliwell, ``Scalar
Fields In Cosmology With An Exponential Potential,'' Phys.\ Lett.\
B {\bf 185} (1987) 341.
}
\lref\ColeyMJ{
A.~A.~Coley and R.~J.~van den Hoogen,
``The dynamics of multi-scalar field cosmological models and assisted
inflation,''
Phys.\ Rev.\ D {\bf 62}, 023517 (2000)
[arXiv:gr-qc/9911075].
}
\lref\BillyardHV{
A.~P.~Billyard, A.~A.~Coley and R.~J.~van den Hoogen,
``The stability of cosmological scaling solutions,''
Phys.\ Rev.\ D {\bf 58}, 123501 (1998)
[arXiv:gr-qc/9805085].
}
\lref\LiddleTB{
A.~R.~Liddle,
``Power Law Inflation With Exponential Potentials,''
Phys.\ Lett.\ B {\bf 220}, 502 (1989).
}
\lref\ChaudhuriEE{
S.~Chaudhuri and D.~A.~Lowe,
``Monstrous String-String Duality,''
Nucl.\ Phys.\ B {\bf 469}, 21 (1996)
[arXiv:hep-th/9512226].
}
\lref\Staro{A.A. Starobinsky, in: {\it Fundamental Interactions} (MGPI
Press, Moscow, 1984), p. 55.}
\lref\Lbook{A.D. Linde, {\sl Particle Physics and Inflationary Cosmology,}
(Harwood, Chur, Switzerland, 1990).}
\lref\LindeXX{
A.~D.~Linde, D.~A.~Linde and A.~Mezhlumian,
``From the Big Bang theory to the theory of a stationary universe,''
Phys.\ Rev.\ D {\bf 49}, 1783 (1994)
[arXiv:gr-qc/9306035].
}
\lref\MijicUD{
M.~Mijic,
``Stochastic Inflation And Quantum Cosmology,''
Int.\ J.\ Mod.\ Phys.\ A {\bf 6}, 2685 (1991).
}
\lref\MijicQX{
M.~Mijic,
``Random Walk After The Big Bang,''
Phys.\ Rev.\ D {\bf 42}, 2469 (1990).
}
\lref\DineHE{
M.~Dine and N.~Seiberg,
``Is The Superstring Weakly Coupled?,''
Phys.\ Lett.\ B {\bf 162}, 299 (1985).
}
\lref\BeckerGJ{
K.~Becker and M.~Becker,
``M-Theory on Eight-Manifolds,''
Nucl.\ Phys.\ B {\bf 477}, 155 (1996)
[arXiv:hep-th/9605053].
}
\lref\BeckerPM{
K.~Becker and M.~Becker,
``Supersymmetry breaking, M-theory and fluxes,''
JHEP {\bf 0107}, 038 (2001)
[arXiv:hep-th/0107044].
}
\lref\BBHL{ K.~Becker, M.~Becker, M.~Haack and J.~Louis,
``Supersymmetry breaking and $\alpha'$-corrections to flux induced
potentials,'' JHEP {\bf 0206}, 060 (2002) [arXiv:hep-th/0204254].
}
\lref\MichelsonPN{
J.~Michelson,
 ``Compactifications of type IIB strings to four dimensions with  non-trivial
classical potential,''
Nucl.\ Phys.\ B {\bf 495}, 127 (1997)
[arXiv:hep-th/9610151].
}
\lref\HaackJZ{
M.~Haack and J.~Louis,
``M-theory compactified on Calabi-Yau fourfolds with background flux,''
Phys.\ Lett.\ B {\bf 507}, 296 (2001)
[arXiv:hep-th/0103068].
}
\lref\LouisNY{
J.~Louis and A.~Micu,
 ``Type II theories compactified on Calabi-Yau threefolds in the presence  of
background fluxes,''
Nucl.\ Phys.\ B {\bf 635}, 395 (2002)
[arXiv:hep-th/0202168].
}
\lref\DineFW{
M.~Dine,
``Is there a string theory landscape: Some cautionary notes,''
arXiv:hep-th/0402101.
}
\lref\GukovYA{
S.~Gukov, C.~Vafa and E.~Witten,
``CFT's from Calabi-Yau four-folds,''
Nucl.\ Phys.\ B {\bf 584}, 69 (2000)
[Erratum-ibid.\ B {\bf 608}, 477 (2001)]
[arXiv:hep-th/9906070].
}
\lref\DouglasUM{
M.~R.~Douglas,
``The statistics of string / M theory vacua,''
JHEP {\bf 0305}, 046 (2003)
[arXiv:hep-th/0303194].
}
\lref\DasguptaSS{
K.~Dasgupta, G.~Rajesh and S.~Sethi,
``M theory, orientifolds and G-flux,''
JHEP {\bf 9908}, 023 (1999)
[arXiv:hep-th/9908088].
}
\lref\GreeneGH{
B.~R.~Greene, K.~Schalm and G.~Shiu,
``Warped compactifications in M and F theory,''
Nucl.\ Phys.\ B {\bf 584}, 480 (2000)
[arXiv:hep-th/0004103].
}
\lref\CurioSC{
G.~Curio, A.~Klemm, D.~Lust and S.~Theisen,
 ``On the vacuum structure of type II string compactifications on  Calabi-Yau
spaces with H-fluxes,''
Nucl.\ Phys.\ B {\bf 609}, 3 (2001)
[arXiv:hep-th/0012213].
}
\lref\AshokGK{
S.~Ashok and M.~R.~Douglas,
``Counting flux vacua,''
JHEP {\bf 0401}, 060 (2004)
[arXiv:hep-th/0307049].
}
\lref\DouglasKP{
M.~R.~Douglas,
``Statistics of string vacua,''
arXiv:hep-ph/0401004.
}
\lref\Escoda{C.~Escoda, M.~Gomez-Reino and F.~Quevedo,
``Saltatory de Sitter string vacua,''
JHEP {\bf 0311} (2003) 065 [arXiv:hep-th/0307160].
}
\lref\BoussoXA{
R.~Bousso and J.~Polchinski,
 ``Quantization of four-form fluxes and dynamical neutralization of the
cosmological constant,''
JHEP {\bf 0006}, 006 (2000)
[arXiv:hep-th/0004134].
}
\lref\KachruAW{
S.~Kachru, R.~Kallosh, A.~Linde and S.~P.~Trivedi,
``De Sitter vacua in string theory,''
Phys.\ Rev.\ D {\bf 68}, 046005 (2003)
[arXiv:hep-th/0301240].
}
\lref\ColemanAW{
S.~R.~Coleman and F.~De Luccia,
``Gravitational Effects On And Of Vacuum Decay,''
Phys.\ Rev.\ D {\bf 21}, 3305 (1980).
}
\lref\heretics{
T.~Banks,
``Heretics of the false vacuum: Gravitational effects on and of vacuum decay.
II,'' arXiv:hep-th/0211160.
}
\lref\GuthPN{
A.~H.~Guth and E.~J.~Weinberg,
``Could The Universe Have Recovered From A Slow First Order Phase
Transition?,''
Nucl.\ Phys.\ B {\bf 212}, 321 (1983).
}
\lref\HawkingGA{
S.~W.~Hawking, I.~G.~Moss and J.~M.~Stewart,
``Bubble Collisions In The Very Early Universe,''
Phys.\ Rev.\ D {\bf 26}, 2681 (1982).
}
\lref\HawkingFZ{
S.~W.~Hawking and I.~G.~Moss,
``Supercooled Phase Transitions In The Very Early Universe,''
Phys.\ Lett.\ B {\bf 110}, 35 (1982).
}
\lref\JarvUK{
L.~Jarv, T.~Mohaupt and F.~Saueressig,
``Quintessence cosmologies with a double exponential potential,''
arXiv:hep-th/0403063.
}
\lref\ReyZK{
S.~J.~Rey,
``Dynamics Of Inflationary Phase Transition,''
Nucl.\ Phys.\ B {\bf 284}, 706 (1987).
}
\lref\LindeGS{
A.~D.~Linde,
``Quantum creation of an open inflationary universe,''
Phys.\ Rev.\ D {\bf 58}, 083514 (1998)
[arXiv:gr-qc/9802038].
}
\lref\WohlfarthKW{
M.~N.~R.~Wohlfarth,
``Inflationary cosmologies from compactification,''
Phys.\ Rev.\ D {\bf 69}, 066002 (2004)
[arXiv:hep-th/0307179].
}
\lref\DysonNT{
L.~Dyson, J.~Lindesay and L.~Susskind,
``Is there really a de Sitter/CFT duality,''
JHEP {\bf 0208}, 045 (2002)
[arXiv:hep-th/0202163].
}
\lref\SusskindKW{
L.~Susskind,
``The anthropic landscape of string theory,''
arXiv:hep-th/0302219.
}
\lref\BanksES{
T.~Banks, M.~Dine and E.~Gorbatov,
``Is there a string theory landscape?,''
arXiv:hep-th/0309170.
}
\lref\SpergelCB{
D.~N.~Spergel {\it et al.},
 ``First Year Wilkinson Microwave Anisotropy Probe (WMAP) Observations:
Determination of Cosmological Parameters,''
Astrophys.\ J.\ Suppl.\  {\bf 148}, 175 (2003)
[arXiv:astro-ph/0302209].
}
\lref\GiddingsZW{
S.~B.~Giddings,
``The fate of four dimensions,''
Phys.\ Rev.\ D {\bf 68}, 026006 (2003)
[arXiv:hep-th/0303031].
}
\lref\GKP{
S.~B.~Giddings, S.~Kachru and J.~Polchinski,
``Hierarchies from fluxes in string compactifications,''
Phys.\ Rev.\ D {\bf 66}, 106006 (2002)
[arXiv:hep-th/0105097].
}
\lref\BeckerGW{
M.~Becker, G.~Curio and A.~Krause,
``De Sitter vacua from heterotic M-theory,''
arXiv:hep-th/0403027.
}
\lref\CopelandET{
E.~J.~Copeland, A.~R.~Liddle and D.~Wands,
``Exponential potentials and cosmological scaling solutions,''
Phys.\ Rev.\ D {\bf 57}, 4686 (1998)
[arXiv:gr-qc/9711068].
}
\lref\SaltmanSN{
A.~Saltman and E.~Silverstein,
``The scaling of the no-scale potential and de Sitter model building,''
arXiv:hep-th/0402135.
}
\lref\RatraRM{
B.~Ratra and P.~J.~Peebles,
``Cosmological Consequences Of A Rolling Homogeneous Scalar Field,''
Phys.\ Rev.\ D {\bf 37}, 3406 (1988).
}
\lref\TownsendRV{
For a review, see: P.~K.~Townsend,
``Cosmic acceleration and M-theory,''
arXiv:hep-th/0308149.
}
\lref\Townsend{
P.~K.~Townsend and M.~N.~R.~Wohlfarth,
``Accelerating cosmologies from compactification,''
Phys.\ Rev.\ Lett.\  {\bf 91}, 061302 (2003)
[arXiv:hep-th/0303097].
}
\lref\Emparan{
R.~Emparan and J.~Garriga,
``A note on accelerating cosmologies from compactifications and S-branes,''
JHEP {\bf 0305}, 028 (2003)
[arXiv:hep-th/0304124].
}
\lref\Cornalba{
L.~Cornalba and M.~S.~Costa,
``A new cosmological scenario in string theory,''
Phys.\ Rev.\ D {\bf 66}, 066001 (2002)
[arXiv:hep-th/0203031].
}
\lref\OhtaA{
N.~Ohta,
``Accelerating cosmologies from S-branes,''
Phys.\ Rev.\ Lett.\  {\bf 91}, 061303 (2003)
[arXiv:hep-th/0303238].
}
\lref\Roy{
S.~Roy,
``Accelerating cosmologies from M/string theory compactifications,''
Phys.\ Lett.\ B {\bf 567}, 322 (2003)
[arXiv:hep-th/0304084].
}
\lref\Wohlfarth{
M.~N.~R.~Wohlfarth,
``Accelerating cosmologies and a phase transition in M-theory,''
Phys.\ Lett.\ B {\bf 563}, 1 (2003)
[arXiv:hep-th/0304089].
}
\lref\Chen{
C.~M.~Chen, P.~M.~Ho, I.~P.~Neupane and J.~E.~Wang,
``A note on acceleration from product space compactification,''
JHEP {\bf 0307}, 017 (2003)
[arXiv:hep-th/0304177].
}
\lref\accel{
C.~M.~Chen, P.~M.~Ho, I.~P.~Neupane, N.~Ohta and J.~E.~Wang,
``Hyperbolic space cosmologies,''
JHEP {\bf 0310}, 058 (2003)
[arXiv:hep-th/0306291].
}
\lref\OhtaB{ N.~Ohta, ``A study of accelerating cosmologies from
superstring/M theories,'' Prog.\ Theor.\ Phys.\  {\bf 110}, 269
(2003) [arXiv:hep-th/0304172].
}
\lref\Vieira{
P.~G.~Vieira,
``Late-time cosmic dynamics from M-theory,''
arXiv:hep-th/0311173.
}
\lref\Bergshoeff{
E.~Bergshoeff, A.~Collinucci, U.~Gran, M.~Nielsen and D.~Roest,
``Transient quin\-tess\-ence from group manifold reductions or how
all roads lead to Rome,''
Class.\ Quant.\ Grav.\  {\bf 21}, 1947 (2004)
[arXiv:hep-th/0312102].
}
\lref\Russo{
J.~G.~Russo,
``Exact solution of scalar-tensor cosmology with exponential
potentials and transient acceleration,''
arXiv:hep-th/0403010.
}
\lref\NeupaneCS{
I.~P.~Neupane,
``Accelerating cosmologies from exponential potentials,''
arXiv:hep-th/0311071.
}
\lref\leblond{
F.~Leblond, D.~Marolf and R.C.~Myers,
``Tall tales from de Sitter space. I: Renormalization group flows,''
JHEP {\bf 0206}, 052 (2002)
[arXiv:hep-th/0202094].
}
\lref\Bousso{
R.~Bousso, O.~DeWolfe and R.C.~Myers,
``Unbounded entropy in spacetimes with positive cosmological constant,''
Found.\ Phys.\  {\bf 33}, 297 (2003)
[arXiv:hep-th/0205080].
}

\Title{\vbox{\baselineskip12pt
\hbox{hep-th/0404220}
}}
{\vbox{\centerline{Spontaneous decompactification}
}}
\centerline{{\ticp Steven B. Giddings}\footnote{$^\dagger$}
{Email address:
giddings@physics.ucsb.edu} }
\centerline{ {\sl Department of Physics\footnote{$^*$}{Current address.}} and
{\sl Kavli Institute for Theoretical Physics}}
\centerline{\sl University of California}
\centerline{\sl Santa Barbara, CA 93106-9530}
\bigskip
\centerline{and}
\centerline{{\ticp Robert C. Myers}\footnote{$^\ddagger$}
{Email address: rmyers@perimeterinstitute.ca
} }
\centerline{\sl Perimeter Institute for Theoretical Physics}
\centerline{\sl 35 King Street North, Waterloo, Ontario N2J 2W9, Canada}
\centerline{and}
\centerline{\sl Department of Physics, University of Waterloo}
\centerline{\sl Waterloo, Ontario N2L 3G1, Canada}
\bigskip
\centerline{\bf Abstract} Positive vacuum energy together with
extra dimensions  of space imply that our four-dimensional
Universe is unstable, generically to decompactification of the
extra dimensions.  Either quantum tunneling or thermal
fluctuations carry one past a barrier into the decompactifying
regime.  We give an overview of this process, and examine the
subsequent expansion into the higher-dimensional geometry.  This
is governed by certain fixed-point solutions of the evolution
equations, which are studied for both positive and negative
spatial curvature.  In the case where there is a
higher-dimensional cosmological constant, we also outline a
possible mechanism for compactification to a four-dimensional de
Sitter cosmology.

\Date{}

\newsec{Introduction}

We now have good reason to believe that  we live in an
accelerating Universe; this point has particularly been brought
home with the recent WMAP results\SpergelCB, combined with earlier
cosmological observations.  It is also widely believed that our
fundamental description of nature should involve extra small
dimensions of space.  These two statements alone lead one to a
very general argument that the Universe as we know it is unstable
to a  catastrophic transition\GiddingsZW.

The generic instability is for the  extra dimensions of space to
begin to grow, and our world to evolve into a higher-dimensional
one.  However, depending on the details of the potential for the
shape and size moduli of the extra dimensions, there may also be
basins of attraction with negative potential, which lead to
equally catastrophic big crunches.  The instability towards
expansion of the extra dimensions was first argued for in the
context of string theory by Dine and Seiberg\DineHE.  Their
arguments were based on supersymmetry.  However, \GiddingsZW\
points out that the underlying mechanism is a simple dynamical
one, driven by the dynamics of long-distance gravity, and is
independent of the existence of supersymmetry.

Examples of potentials exhibiting  this instability are now being
widely studied in string theory.  Flat directions in moduli space
have been a long-standing problem in string theory. Recent
developments in compactifications with fluxes and
branes\refs{\PolchinskiSM\ChaudhuriEE\BeckerGJ\MichelsonPN\BeckerPM\HaackJZ\LouisNY\GukovYA\DasguptaSS\GreeneGH-\CurioSC}
have provided examples of dynamics that fixes these
moduli\refs{\GKP,\KachruAW}. In particular, \refs{\KachruAW}
showed that by adding an anti-D3 brane to the solutions of
\refs{\GKP}, one can lift the vacuum energy and find locally
stable minima with  positive cosmological constant.  Subsequent
work on other approaches to vacua with positive cosmological
constant has included \refs{\Escoda\SaltmanSN-\BeckerGW}.

These developments have   led to much discussion, as well as
criticism\refs{\BanksES,\DineFW}, of the resulting picture of a
``landscape" of stringy vacua.  This picture has led to a forceful
resurrection\SusskindKW\ of the idea that constants of nature --
particularly the cosmological constant -- are determined
anthropically; the large number of possible fluxes and resulting
vacua\refs{\DouglasUM,\AshokGK} (for a review see \DouglasKP)
together with the observation that in a sufficiently large
distribution one expects to find a small enough cosmological
constant\BoussoXA, have given strong fuel to this possibility.

One of the criticisms from  \BanksES\ is potentially relevant to
our discussion, and so deserves comment.  Banks and Dine argue
that an effective potential description of the landscape is not
justified and cannot be trusted.  Underlying this argument is the
realization that there are regions in the combined field space of
the moduli and metrics where the dynamics becomes strongly
coupled.  While it is certainly not inconceivable that this could
render the entire picture inconsistent, we take a more sanguine
perspective.  It may well be that there are dragons off in the
mountains of the landscape.
However, the valley we find ourselves in seems perfectly tame, and
we expect that dynamics nearby is likewise tame.  Of course we'd
very much like to understand the strongly coupled dynamics to
understand how we arrived to our present vacuum, but for now we'll
take the perspective that the dynamics of the full quantum
wavefunction has somehow deposited us here, and our problem is to
see what happens next. We provisionally accept that the effective
potential is a useful tool in this investigation.

In the next section  we review the derivation of the effective
potential for the radial dilaton modulus.  In particular, this
analysis shows that modular  landscapes have a generic feature,
much like the ``front range" of Colorado -- the mountains taper
off to a semi-infinite plain.  As we roll into this region, the
extra dimensions of space expand.  Section two also discusses
examples of potentials that can be obtained from fluxes and
branes, for example in string theory.  Section three turns to the
problem of analyzing the asymptotic dynamics of solutions that
have escaped a metastable minimum and are running to
infinitely-expanded extra dimensions.  Asymptotically these
solutions become fixed-point solutions, whose form we derive from
the equations of motion. Section four discusses mechanisms for
escape  from the metastable minimum -- thermal excitation over or
tunneling through the barrier. Finally, in section five, we
assemble these results with a general discussion of the
decompactification transition out of (our?) metastable de Sitter
space and into the decompactifying regime, with the resulting
higher-dimensional evolution. We also briefly discuss a scenario
in which transitions can take place back and forth between a
higher-dimensional dS space and the four-dimensional one; in this
case, the dominant configuration should be determined by the
relative entropies, providing a possible mechanism for
compactification.  An appendix provides a more detailed
phase-plane analysis, for both positive and negative spatial
curvatures,  of the solutions arising in our discussion of
decompactification,

Spontaneous decompactification has also been recently  discussed
for simple compactifications with fluxes in \JarvUK, which
appeared while our paper was being written up.

\newsec{Radial dilaton dynamics: generalities}

\subsec{Dimensional reduction}

We begin by discussing  dynamics of the radial dilaton,
following\refs{\GiddingsZW}.  Specifically, suppose that we begin
with ($d$+4)-dimensional action
\eqn\Act{S=\int d^{d+4}X \sqrt{-G}\left[ M_P^{d+2} \cR + \cL(\psi) + {\hat \cL}(\psi, \cR)\right]\ ,}
where $X$ and $G$ are  the coordinates and metric of the full
($d$+4)-dimensional spacetime, $M_P$ is the ($d$+4)-dimensional Planck
mass, $\cR$ is the Ricci scalar, $ \cL(\psi)$ is the lagrangian
representing the leading contribution of generic matter sources in
a derivative expansion, possibly including  localized sources such
as D-branes, and  ${\hat \cL}(\psi,\cR)$ summarizes possible
corrections to the leading lagrangian that involve higher powers
of the curvature and/or higher derivatives of matter fields. This
action may be the effective action for string theory, or for some
other fundamental theory of gravity. For our present purposes we
assume that all the moduli except the overall volume modulus of
the internal space are fixed, \eg\ as in \GKP; we will consider
the coupled dynamics of this modulus and the four-dimensional
metric.  (It has been argued in \WohlfarthKW\  that in the case of
multiple moduli with exponential potentials, a single term
dominates, leading to an obvious generalization of our analysis to
the multi-moduli case.) Thus we assume that \Act\ has solutions of
the form\foot{Here we suppress a possible warp factor, which
should not change our general picture.}
\eqn\rescale{ds^2 = ds_4^2 + R^2(x)g_{mn}(y) dy^m dy^n \ .}

The action governing  solutions with $R(x)=e^{D(x)}$ varying
slowly on the compactification scale follows from dimensional
reduction of \Act. The Einstein-Hilbert term gives a 4d effective
action
\eqn\ehred{S_{EH}= M_P^{d+2}  V_d \int d^4x \sqrt{-g_4} \left[
e^{dD(x)} \cR_4 + d(d-1) (\nabla D)^2 e^{dD(x)} + e^{(d-2)D}
\cR_d\right]\ ,}
where $V_d$ and $\cR_d$ are  the volume and curvature of the
$d$-dimensional compact metric $g_{mn}$.  However, in this frame
the 4d effective Planck mass varies with $R$.  To make contact
with usual treatments of 4d dynamics, one should choose a new set
of units by performing the Weyl rescaling
\eqn\weylr{g_{4\mu\nu} \rightarrow e^{-dD} g_{4\mu\nu}\ ;}
the 4d Planck mass then becomes
\eqn\mfour{M_4^2 = M_P^{d+2} V_d\ .}
The additional terms in  \Act\ then contribute an additional
potential for $D$.  The net result is an action of the form
\eqn\redact{ S = M_4^2\int  d^4 x \sqrt{-g_4} \Biggl\{  \cR_4 -
{1\over 2}  d(d+2)(\nabla D)^2 - V(D) \Biggr\}\ .}

A central observation of  \GiddingsZW\ is that any physics that
stabilizes the radial dilaton from runaway to infinite volume,
$D\rightarrow\infty$, must do so only locally -- the Weyl
rescaling \weylr\ introduces an inverse power of the
volume-squared into the potential.  This means that  to stabilize
the radial dilaton, the higher-dimensional dynamics would have to
produce an energy {\it density} growing at least as fast as the
internal {\it volume} for large volume.  This does not appear
realizable in any realistic physical theory.

\Ifig{\Fig\figone}{A sketch  of a potential with a metastable de
Sitter region, and runaway to infinite dilaton.}{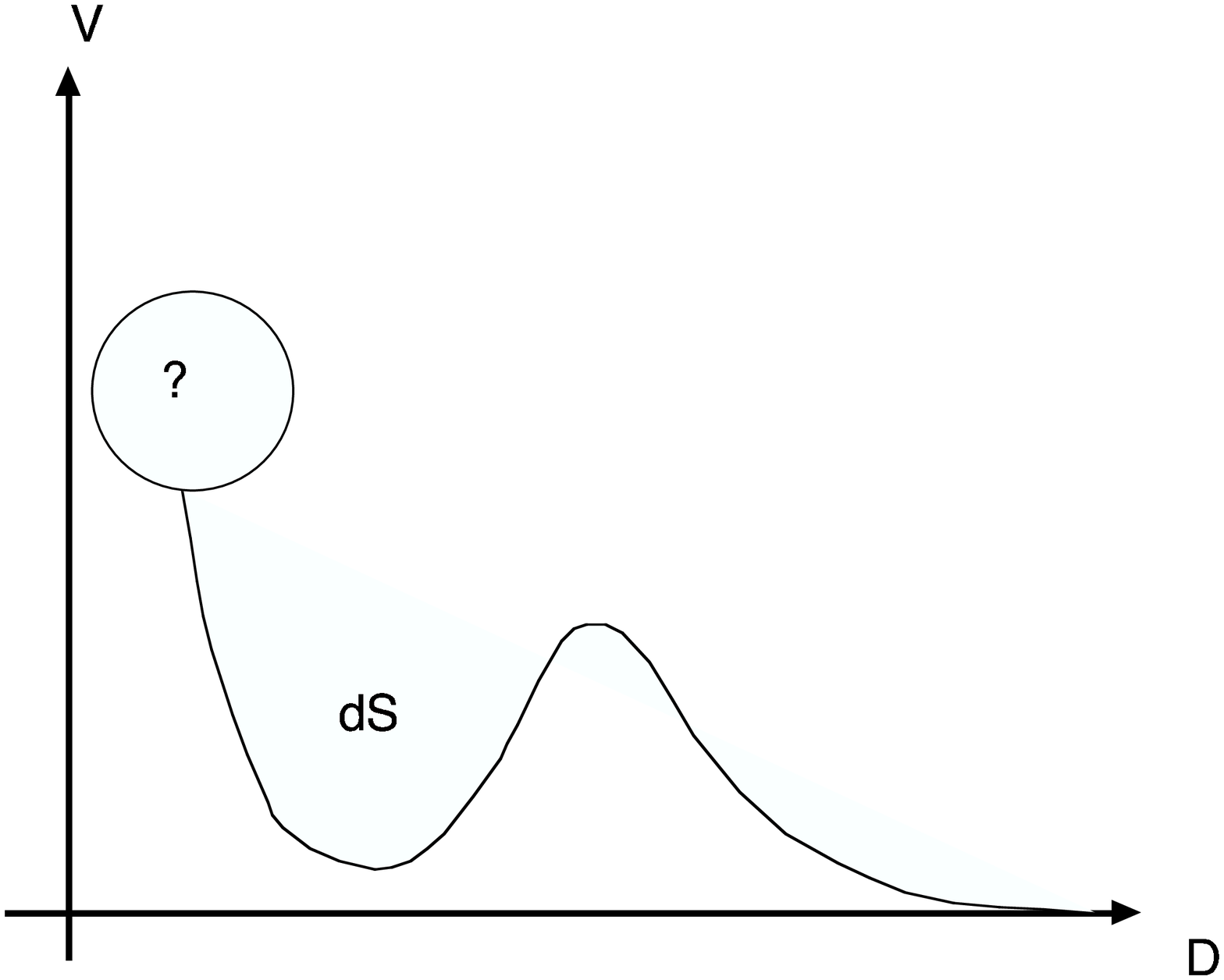}{4}

\Ifig{\Fig\figtwo}{A sketch  of a potential with a metastable de
Sitter region, and a minimum with negative cosmological constant.
If the Universe fluctuates into the AdS basin of attraction, it
evolves to a big crunch singularity.}{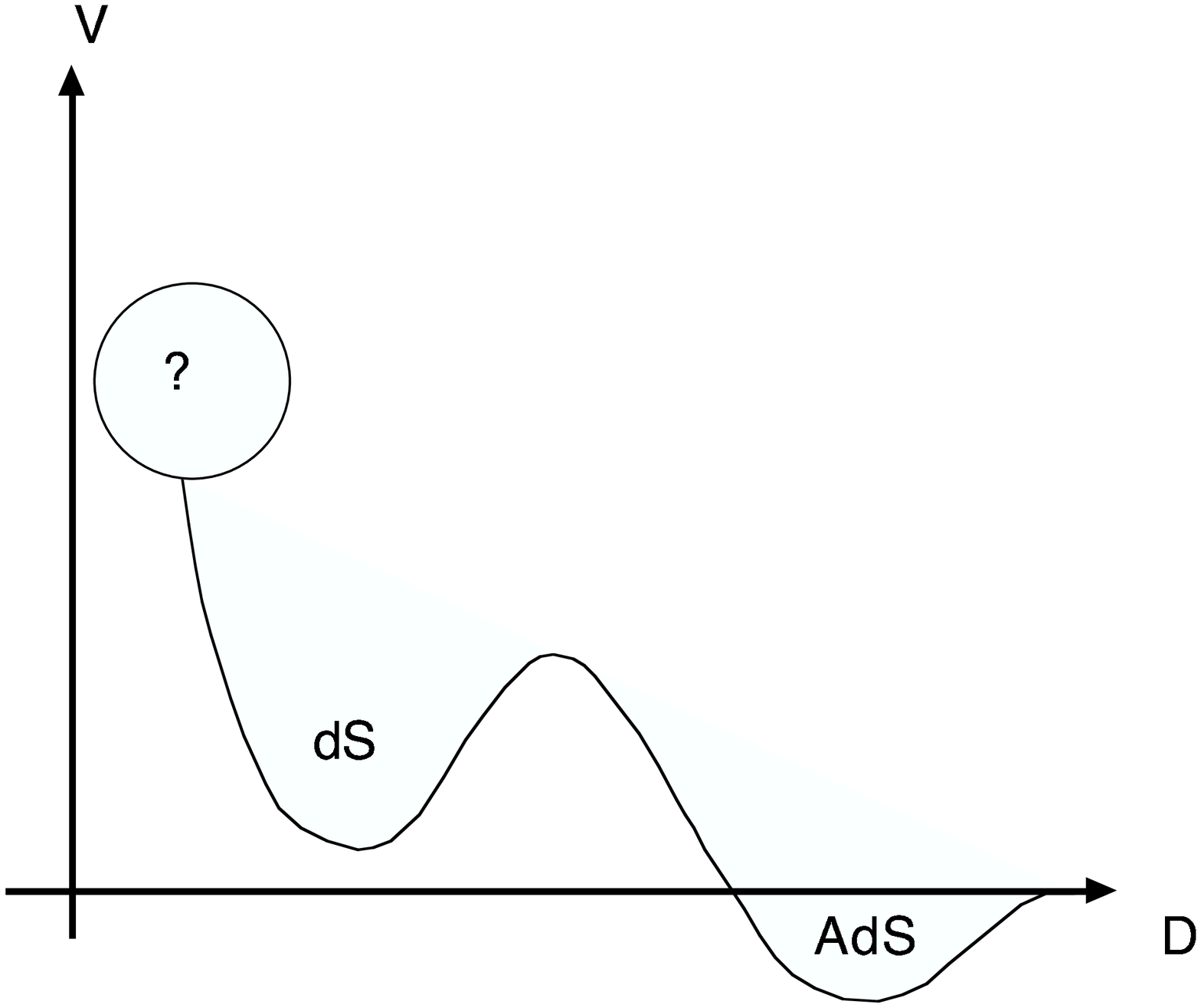}{4}

Therefore if there is a  positive potential minimum representing
the present de Sitter phase of our Universe, this minimum must be
metastable, and the potential should generically appear as in
fig.~1 or fig.~2.  Fig.~2 produces an instability to a big crunch
spacetime\refs{\ColemanAW,\heretics}.  Our focus will be the more
generic case of runaway to infinitely extended extra dimensions,
as follows from fig.~1.

Of course, in full generality  there will be a nontrivial
multi-dimensional moduli space of compact manifolds, and a
potential function on this space.   This has recently been
explored in the case of string theory for the region of string
configuration space corresponding to flux compactifications, and
the resulting configuration space dubbed\SusskindKW\ `the landscape.'  The
argument given in \GiddingsZW\  implies that the landscape has a
general feature similar to the geology of the front range of the
Rocky Mountains in North America:  the mountains and valleys of
the landscape roll off into a flat plain, extending to infinity.

\subsec{The dilaton potential -- examples}

In general the dilaton potential  receives contributions from
non-trivial field configurations in the extra dimensions.  For
illustration, it's particularly fruitful to consider examples
provided from string theory, particularly branes and fluxes.
There may well also be other effects, both perturbative and
non-perturbative in the couplings.

Some of the examples from string  theory were enumerated in
\GiddingsZW.  A $q$-form flux field with action
\eqn\formact{S_p \propto - \int d^{d+4} X \sqrt{-G} {F_q^2\over q!}\ ,}
gives a potential
\eqn\fluxpot{V_F\propto R^{-d-2q} \ }
and a space-filling $p$-brane with action
\eqn\braneact{S_p =-{\mu_p\over g_s} \int_{M_4\times \cC_{p-3} } dV_{p+1}\ ,}
gives a potential
\eqn\branepot{ V_p \propto R^{p-3-2d }\ . }
{}From the action \ehred, curvature on the internal
manifold can give a potential
\eqn\curvpot{ V_R\propto R^{-d-2}\ .   } Nonperturbative effects
give other contributions,
\eg\ \refs{\KachruAW} included a potential of the form
\eqn\nonpertV{V= Be^{-2aR^4}/R^s\ ,}
with different constants $a$ and $s$ arising  from euclidean D3 brane
instantons or gluino condensation.  Finally $\alpha'$ corrections
were argued in \refs{\BBHL} to produce potential contributions of
the form
\eqn\beckercor{V_K\propto {1\over R^{18}}\ .}

\newsec{Asymptotic dilaton dynamics}

\subsec{Fixed-point dynamics}

Quantitative details of the de Sitter  decay will depend on the
contribution of these and other possible terms to the potential.
However, we'll be particularly interested in the asymptotic
structure of the expanding region of higher-dimensional space.
The dynamics inside this region is governed by the asymptotics of
the potential at large dilaton, and for this in general we need
only to focus on the leading term.  Since this will typically be
of the form $V\propto R^n$, let us analyze more closely the
dilaton dynamics with a single such term in the potential.   With
canonically normalized dilaton,
\eqn\normd{\phi=\sqrt{d(d+2)} D\ ,}
we find an action of the form
\eqn\newact{S=M_4^2\int d^4x \sqrt{-g}  \left[\cR_4 -\hf (\nabla
\phi)^2 - V(\phi)\right]\ }
with
\eqn\exppot{V(\phi)={A\over 2} e^{-\alpha \phi}\ .}
In these conventions,
\eqn\alphaex{ \eqalign{\alpha &=\sqrt{d\over d+2} \left(1+{2q\over d}\right)\quad\quad\quad
        \quad {\rm q-flux}\cr
        &= \sqrt{d\over d+2}\left(1+{d+3-p\over d}\right) \quad {\rm p-brane}\cr
        &= \sqrt{d+2\over d}\qquad\qquad\qquad
        \quad\quad\ \ {\rm internal\ curvature} \ .}}

Friedman-Robertson-Walker solutions for  actions of this form,
with exponential potentials, have been investigated previously in
the literature, see \eg\ \refs{\BBL\BarrowIA\Halliwell\RatraRM\LiddleTB\CopelandET\BillyardHV-\ColeyMJ}.
Recently, there has been a great deal of discussion of analogous solutions
in string- and M-theory, see \eg\
\refs{\TownsendRV\Townsend\Emparan\Cornalba\OhtaA\Roy\Wohlfarth\Chen\accel\OhtaB\NeupaneCS\Vieira\Bergshoeff-\Russo,\JarvUK}.
As we'll see in the next section, such FRW solutions, with spatial
curvature $k=0,\pm1$, are useful in studying the
decompactification transition.  These have the form
\eqn\RWmet{ds_4^2=-dt^2 + a^2(t)d\chi^2\ ,\ \phi=\phi(t)\ }
where $d\chi^2$ is the appropriate spatial metric  with curvature
$k=0,\pm1$. The equations of motion take the form
\eqn\constrainteq{ {{\dot a}^2\over a^2} = {\rho_\gamma\over6} + {{\dot \phi}^2 \over 12} + {V\over 6}\ ,}
\eqn\aeqn{ \left({\dot{a}\over a}\right)^{\kern-5pt\raise
.3ex\hbox{.}} + {\gamma \rho_\gamma\over 4} = - {{\dot
\phi}^2\over 4}\ ,}
\eqn\phieqn{ {\ddot\phi} = -3 \left( { {\dot a}\over a} \right) {\dot \phi} -{dV\over d\phi}\ .}
Here we've included the possibility of  a stress tensor with a
barotropic equation of state,
\eqn\barot{p_\gamma= (\gamma-1)\rho_\gamma \ ,}
for which the density evolves as
\eqn\rhoevol{\dot{\rho}_\gamma=-3\gamma{\dot{a}\over a}\rho_\gamma\ .}
In the case $k=\pm1$, the dominant  effective contribution to
$\rho$ at large $a$ is that summarizing the spatial curvature,
which gives
\eqn\curvr{\rho_{2/3} = -{6k\over a^2}\ .}
Massive matter ($\gamma=1$) or  radiation ($\gamma = 4/3$) will
make subdominant contributions as $a\rightarrow\infty$ in these
cases, but if $k=0$ their contribution can be relevant to this
asymptotic behavior.

The asymptotic behavior of  solutions of these equations is
typically governed by fixed point ``tracker" solutions.  The fixed
points relevant to the cases $k=0,-1$ have been found via phase
space analyses in \refs{\RatraRM,\CopelandET}. A simple extension
of this work covers the case $k=+1$ as well, as described in the
appendix. Let's understand how these solutions arise. The starting
point is the observation that for a scale factor that increases as
a power of time (which will be justified shortly),
\eqn\ascale{a=a_0(ct)^\beta\ ,}
and for an exponential  potential \exppot, equation \phieqn\
drives $e^\phi$ to infinity as a power:
\eqn\phisoln{e^{\alpha\phi} = (ct)^{2}}
where the constant $c$ is fixed to be
\eqn\cparam{c={\alpha\over 2} \sqrt{ A\over 3\beta-1}\ .}
Note that \phieqn\ can  also be integrated to give the
$\phi$-dependent terms in the Freedman equation:
\eqn\phif{ {{\dot \phi}^2 \over 12} + {V\over 6} = {\beta\over \alpha^2 t^2}\ .}
Thus, using \rhoevol, the Freedman equation \constrainteq\ becomes
\eqn\newfreed{ {{\dot a}^2\over a^2} = {c'\over t^{3\gamma\beta}} + {\beta\over \alpha^2 t^2}\ }
for some constant $c'$. First consider the cases $k=0,-1$, where
$c'$ is positive. Which term gives the dominant fixed-point
asymptotic behavior as $t\rightarrow\infty$ depends on the
relative magnitudes of $\alpha$ and $\gamma$.  A phase plane
analysis of the dynamics appears in the appendix, but the basic
features can be understood straightforwardly directly from the
equations.  For $3\gamma\beta>2$, the first term becomes
subdominant as $t\rightarrow\infty$.  Then, using the power law
form of the solution, \ascale, we find $\beta=1/\alpha^2$.  This
means the condition for this fixed point to dominate is
$\alpha^2<3\gamma/2$.  In contrast, for $\alpha^2>3\gamma/2$ both
terms are relevant.  Correspondingly, the power law form \ascale\
now implies $\beta=2/3\gamma$.
The complete phase plane analysis (see \refs{\RatraRM,\CopelandET}
and the appendix) shows that these are indeed the fixed points to
which the generic solution is attracted
--- in the appendix, fixed point (a) for $\alpha^2<3\gamma/2$ and
(b) for $\alpha^2>3\gamma/2$ (see fig. 4). Any other fixed points
are either unstable nodes or saddle points.

The case of positive spatial curvature $k=+1$ allows negative $c'$
and is somewhat more subtle.  For $\alpha^2<3\gamma/2=1$,
depending on initial conditions, the solution may reach a turning
point and reflect to a collapsing universe, $\dot a<0$, rather
than reaching the asymptotic $a\rightarrow\infty$ region.  If this
happens, other terms in the Freedman equation \constrainteq, {\it
e.g.,} due to matter/radiation fields, become important in the
dynamics, and a general analysis is not possible.  However, if the
solution asymptotically  expands, the analysis is similar to the
$k=-1$ case.  For $\alpha^2>1$, solutions generically reach a
turning point and recollapse; this behavior is not well-described
by the power law form \ascale. These features are well illustrated
by the corresponding phase plot in the appendix --- see fig. 5.

We can summarize the resulting asymptotically expanding solutions by
\eqn\aexp{ds^2 = -dt^2 + a_0^2\left[c(t-t_0)\right]^{2\beta}d\chi^2\ ,}
and
\eqn\Dexp{e^\phi =  \left[c(t-t_0)\right]^{2/ \alpha}  \ , }
with $c$ given in \cparam. The constant $a_0$ may be fixed by
equation \constrainteq\ for $k=\pm1$ and the expansion exponent
$\beta$ is given by
\eqn\betadef{\beta = {\rm max}\left({2\over 3\gamma},{1\over \alpha^2}\right)\ .}

\subsec{Specific cases}

Consider first the  case of nonzero spatial curvature, $k=\pm1$.
At large radius, the curvature dominates over the matter
contributions to the Friedmann equations, and curvature
corresponds to $\gamma=2/3$.  Thus for $\alpha>1$, which  is
generically the case for string-theory induced potentials, for
$k=-1$ we have $\beta=1$, giving a non-accelerating solution with
a linearly growing scale factor\foot{However, solutions
approaching this fixed point may be eternally
accelerating\refs{\TownsendRV,\accel,\NeupaneCS}.}; for $k=+1$ we have
collapse. For example, consider a flux-induced potential. We can
easily see from \alphaex\ that
\eqn\alphafl{\alpha_{\rm q-flux}^2= \left(1+2{q\over d}\right)\left(1+2{q-1\over d+2}\right)\ }
which gives $\alpha^2>1$. Similarly for internal curvature, we find
\eqn\alphacurv{\alpha_{\rm curvature}^2 = 1+{2\over d}\ }
so again $\alpha^2>1$.
The case of a brane-induced potential gives, from \alphaex,
\eqn\alphabr{\alpha_{\rm p-brane}^2 = \left(1+{d+3-p\over d}\right)
\left(1+{d+1-p\over d+2}\right)\ .}
For $p\leq d+2$, \ie\ a brane  of codimension one or higher, we
therefore also find $\alpha^2>1$. An exceptional case is that  of
a completely space-filling brane, $p=d+3$, or equivalently
higher-dimensional cosmological constant, which gives
\eqn\alphasp{\alpha=\sqrt{d\over d+2}<1\ .}
The corresponding solution  then has $\beta=1/\alpha^2 = 1+2/d$
and is accelerating.

Other string theory  potentials typically fall off even more
rapidly at infinity, removing them further from this accelerating
case.

Next consider spatially  flat universes, $k=0$.  In these cases,
the asymptotic density of matter or radiation, and in particular
the dominant value of $\gamma$,  is relevant for determining which
attractor governs the dynamics.  Note, however, that  one still
only achieves acceleration for $\alpha<1$.

We will discuss the  $D$-dimensional interpretation of these
solutions in the following section.

\newsec{The decompactification transition}

\subsec{Escaping inflation}

We next turn to the  study of decay of a metastable de Sitter
minimum in the potential for the radial dilaton.  In particular,
we assume that the relevant dynamics for our discussion is that of
the radial modulus, and that other possible moduli of the internal
manifold are fixed, although as we've pointed out this analysis
should extend to the multi-moduli case where that is dominated by
an exponential potential due to a single modulus, as in
\WohlfarthKW.  Thus we consider the lagrangian \newact, where
$V(\phi)$ has a form sketched in fig.~1.
For simplicity, we work in units where $M_4=1$ in the following.

Any given region of  the de Sitter universe corresponding to the
metastable minimum will ultimately decay to a decompactifying
universe.  The time scale depends on the parameters of the
potential.\foot{For a more detailed review of parameters and
dynamics see \refs{\KachruAW}.}  If the minimum corresponds to our
presently inflating phase, we should have $V_0=V(\phi_0)\sim
10^{-120}\ll1$.  For a general potential, without any further fine
tuning, we expect that the height of maximum of the barrier at
$\phi_1$ is $V_1\sim 1$ and that the width of the barrier is also
$\Delta\phi \sim 1$.  There are then two basic mechanisms for
inducing spontaneous decompactification.  The first is for thermal
excitations of de Sitter space to take one over the top of the
barrier.  The second is quantum tunneling through the barrier, as
described by Coleman and de Luccia \refs{\ColemanAW}.

The thermal activation  rate is given by the action of the
Hawking-Moss instanton\refs{\HawkingFZ}, as has been argued from
the stochastic approach to
inflation\refs{\Staro\Lbook\ReyZK\MijicQX\MijicUD\LindeXX-\LindeGS}.
This gives  the probability of a thermal fluctuation to take one
to the top of the barrier.  If the change in entropy for this
fluctuation is $\Delta {\bf S}$, the probability is given by the
usual formula $P=\exp\{\Delta {\bf S}\}$. This probability is
\eqn\hmprob{P_{\rm thermal} = e^{S(\phi_0)-S(\phi_1)}\ ,}
where where $S(\phi_0)=-24\pi^2/V_0$ is the action of the de
Sitter instanton, ie, the four-sphere solution, with
$\phi=\phi_0$, the metastable minimum. Similarly, $S(\phi_1)$ is
the solution with $\phi=\phi_1$, the maximum of the barrier. The de Sitter
recurrence time\refs{\DysonNT} is given by
$T_r=\exp\{S(\phi_0)\}$.  The thermal decay lifetime is thus
shorter than this by an exponential factor:
\eqn\thermt{\tau_{\rm thermal} = e^{-24\pi^2/V_1} T_r\ .}

The tunneling probability is given by\refs{\ColemanAW,\KachruAW}
\eqn\tunprob{P_{\rm tunnel} = \exp\left\{{S(\phi_0)\over [1+(4V_0/3T^2)]^2}\right\} }
where the tension of the bubble wall is
\eqn\tension{T=\int_{\phi_0}^\infty d\phi \sqrt{2V(\phi)}\ .}
For the parameters that we  generically expect, we see that
$V_0\ll T^2$.  The probability \tunprob\ therefore yields a decay
time
%
\eqn\tunnelt{\tau_{\rm tunnel} \sim e^{-{64\pi^2/ T^2}}  T_r\ .}
Comparing \thermt\ and \tunnelt\  shows that which process is
dominant clearly depends on the relative magnitudes of the tension
$T$ and the barrier height $V_1$.

Once the field excites past the  barrier, it classically evolves
towards decompactification.  In the case of thermal activation,
one expects a picture where an entire horizon volume thermally
excites over the barrier.  The boundary conditions then start the
solution near the maximum of the potential, and the resulting
classical solution can roll into the decompactification region.
However, fluctuations are expected to be important in this case,
and so one cannot clearly identify the resulting dynamics as a
$k=0,\pm1$ universe.\foot{We thank to A. Linde for a discussion
on this point.}  However, these may serve as rough indicators for
the dynamics, bearing in mind that different regions may have
different spatial curvatures.  One also expects to have some
excited matter fields in the resulting solution; this is important
to avoid certain collapsing solutions in the case of vanishing
spatial curvature, $k=0$, but more generally it seems plausible
that collapse could occur in some regions, and expansion in
others.

\Ifig{\Fig\figthree}{A representation of  the the Coleman
de-Luccia instanton, describing tunneling from a metastable de
Sitter region into a bubble of space in which the extra dimensions
of space are expanding.  The lower half of the diagram is the
euclidean solution of \ColemanAW, for the potential $V(\phi)$.
This matches onto the lorentzian solution pictured in the upper
half of the diagram.  The straight lines correspond to the $a=0$
surface of the resulting expanding $k=-1$ cosmology; the surfaces
above these are surfaces of constant $t$ in this cosmology.  The
boundary of this decompactifying region is a bubble wall (circular
below, hyperbolic above)  which asymptotically expands at the
speed of light into the metastable de Sitter region.}{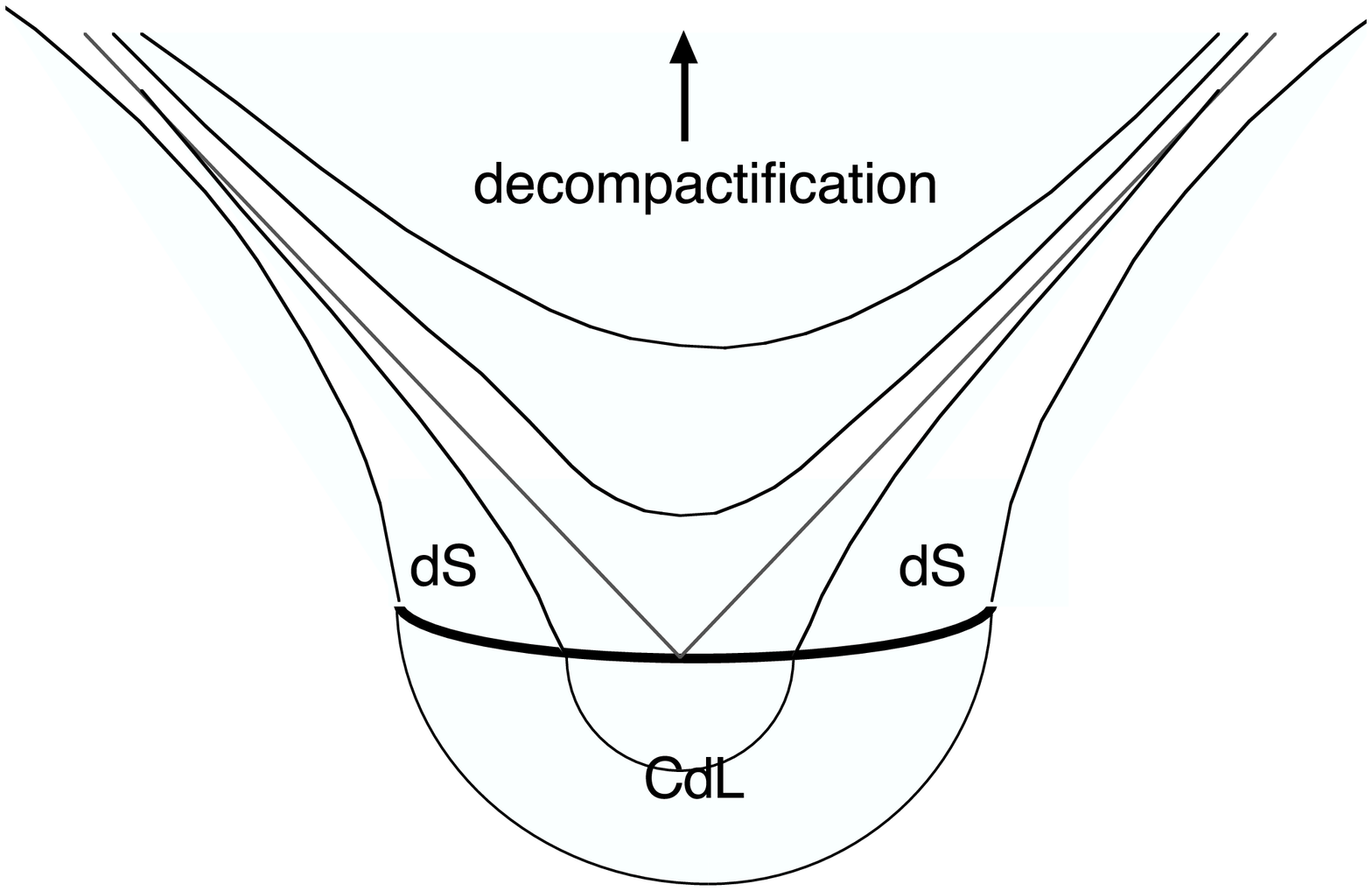}{4}

In the case of tunneling, one can be a  little more specific about
the resulting boundary conditions, by matching the CdL instanton
onto a lorentzian geometry at the turning point.  The $O(4)$
symmetry of the instanton continues to an $SO(3,1)$ symmetry of
the subsequent classical solution,
implying\refs{\ColemanAW,\GuthPN} that one has evolution on
spatial sections with $k=-1$.  These features are illustrated in
fig.~3.  The boundary conditions on the ``initial slice" of the
cosmology are
\eqn\Cdlbc{\dot\phi=0\ ,\ a=0\ ,\ \dot a=1\ {\rm and}\
\phi=\phi_t}
where $\phi_t$ is the value at the turning point.

\subsec{Expansion to higher dimensions}

In either case, even if we know the  potential explicitly it is
typically hard to find an explicit solution.  However, the
fixed-point behavior of these systems tells us that this is not
necessary to understand the asymptotic behavior.  Specifically, we
expect the asymptotic behavior of the decompactifying solution to
be governed by the leading term in the potential as
$a\rightarrow\infty$.  This is expected to generically be one of
the exponentials we've considered, and so decompactification is
asymptotically described by one of the fixed-point solutions of
the preceding section, described by  (in the $k=0,-1$ cases) the solution \aexp-\betadef.

Since, as shown by eqn.~\Dexp,  the compact manifold expands, the
relevant description becomes the higher-dimensional description
rather than that of four dimensions.  Recall that these differ by
the rescaling \weylr, so  in terms of the fundamental units of the
higher-dimensional theory, the asymptotic metric takes the form
\eqn\tendmet{ds_{4+d}^2 = e^{-dD(t)}\left(-dt^2 + a^2(t)
d\chi^2\right) + e^{2D(t)} ds_d^2\ .}
%

In discussing the higher-dimensional  form of these solutions, it
proves easier to work with the coefficient $\la$, related to
$\alpha$ by
\eqn\kdef{ \la=\sqrt{d+2\over d}\alpha\ ;}
thus for fluxes, branes and internal curvature, we find from equation \alphaex\
\eqn\kex{ \eqalign{\la &= 1+{2q\over d}\quad\quad\quad
        \quad {\rm q-flux}\cr
                        &= 1+{d+3-p\over d} \quad {\rm p-brane}\cr
                        &= 1+{2\over d}\qquad\qquad\ \ {\rm internal\ curvature}\ .}}
Then given the asymptotic solution in equations \aexp\ and \Dexp, the higher-dimensional
metric \tendmet\ becomes
\eqn\tendasy{ds_{d+4}^2 = \left(c(t-t_0)\right)^{-2/\la}\left[-dt^2 +
a_0^2\left(c(t-t_0)\right)^{2\beta}d\chi^2\right]
+ \left(c(t-t_0)\right)^{4/\la d} ds_d^2\ .}

For $\la>1$, this metric can be simplified by defining a new time coordinate
\eqn\taudef{\tau ={1\over c}{\la\over \la-1}\left(c(t-t_0)\right)^{1-1/\la}\ ,}
with which we find
\eqn\newmet{ds^2= -{d\tau^2} +a_0^2\left({\hat c}\tau\right)^{2(\beta \la-1)/(\la-1)}d\chi^2
+\left({\hat c} \tau\right)^{4/d(\la-1)}ds_d^2\ .}
Here the constant $\hat c$ is given by
\eqn\cdef{{\hat c}={\la-1\over \la}c\ ;}
recall that $c$ was given in eq.~\cparam, and $\beta$ in eq.~\betadef. 
Both the three large dimensions and  the compact dimensions
expand. However, the compact dimensions expand more slowly unless
\eqn\relexp{(\beta \la-1)d\leq2\ .}
Thus for $k=-1$ they do so in the case of flux-induced potentials
and of potentials arising from branes of codimension greater than
two.

%
The special case of a wrapped $p=d+3$  space-filling brane
corresponds to $\la=1$ and $\beta = 1+2/d$. Now we define
\eqn\infltime{\tau={1\over c} \ln(c(t-t_0))\ ,}
and the metric becomes
\eqn\dsmet{ds^2 = -d\tau^2 + e^{4c\tau/d}\left(a_0^2 d\chi^2 + ds_d^2\right)\ .}
Thus, locally, the metric looks like a  patch of inflating
deSitter space --- for a discussion of such asymmetric foliations
of deSitter space, see \refs{\leblond}.

As the size of the compact dimensions  become comparable to that
of the other three, in general the four-dimensional effective
description will break down and one must describe the solutions as
fully higher-dimensional solutions.  Since the field
configurations, \eg\ branes and fluxes, on the compact
manifold will generically be non-uniform,  this means that the
resulting higher-dimensional solution is generically non-uniform.
Moreover, fluctuations of modes that get light in this limit can
become relevant. Nonetheless, in many cases we expect the
four-dimensional solutions that we have described to lift to give
a higher-dimensional solution, and thus a reasonably accurate
picture of the higher-dimensional dynamics.

\newsec{Overview: the fall}

Let us assemble the pieces of the  picture that we have discussed.
Beginning with our Universe in its currently inflating phase,
ultimately a fluctuation will carry it out of our metastable
minimum.  This may either be a thermal fluctuation over the
barrier, or a tunneling event through the barrier.  In the thermal
case, an entire horizon-sized region fluctuates over the barrier.
In the tunneling case, a bubble forms, and then expands.

The relevant timescales are extremely  long, as they contain a
factor of the recurrence time, $T_r\sim \exp\{10^{120}\}$.  The
decay time, expressed as a fraction of the recurrence time, is
given in eqns.~\thermt, \tunnelt, and depends on the relative
magnitude of the bubble tension and the barrier height.

What happens next depends on which basin  of attraction the
fluctuation takes us into.  There may be accessible basins of
attraction with negative effective cosmological constant; in that
case the Universe will undergo a big crunch.  Alternatively, we
know that there is generically an infinite basin of attraction
corresponding to decompactification.  This is the one that we have
considered in this paper.  We also expect this basic picture to
extend to the case of multi-moduli runaway; ref.~\WohlfarthKW\ has
argued that the case of multiple runaway scalars with exponential
potentials can be reduced to the single-scalar case.  There also
may be basins of attraction where some of the dimensions of space
decompactify and others remain stabilized.  These would be
described by straightforward generalization of our analysis.

Once we find ourselves in such a basin of  attraction, the
Universe decompactifies, with asymptotic dynamics typically given by
\newmet\ or \dsmet.  In the case of a thermal fluctuation, we
expect an entire horizon-sized region to evolve into the region
where higher-dimensional dynamics is relevant, but the actual
configuration may be highly non-uniform, with collapse in some
regions and expansion in others.
It is in this context that we imagine that our results for k=+1
might be applicable. That is, they might describe the dynamics of
a small patch with positive spatial curvature in a larger
inhomogeneous solution. The surprising result from the phase plane
analysis --- see appendix --- is that for typical potentials, \eg\
induced by fluxes or branes, the solutions generically evolve to a
big crunch. We expect that such regions collapse into black holes
and are thus casually disconnected from the expanding
regions.\foot{A similar description has been advocated for the
crunch singularities appearing in transitions towards AdS
minima\refs{\SusskindKW}.}

In the case of tunneling, a bubble of higher-dimensional space
forms, and its walls expand into the four-dimensional region at
the speed of light.  The solution inside the bubble is given by
the asymptotic dynamics that we have described.  As in old
inflation, the majority of space continues to be inflating
four-dimensional space\refs{\GuthPN,\HawkingGA}, but any given
point will ultimately transition into the higher-dimensional
realm.

Now we return to a special case of of space-filling
brane with $p=d+3$, \ie\ a (positive) cosmological constant in ($d$+4)
dimensions. In this case, the landscape as illustrated in fig. 1
is deceptive. Despite the appearance of the usual infinite plain,
there are no solutions rolling towards flat decompactified space
in ($d$+4) dimensions. Rather as we have explicitly shown in \dsmet,
the (apparently) metastable dS space transitions to a spacetime
which asymptotes to ($d$+4)-dimensional de Sitter space. Recall from
the discussion around \alphasp\ that this was the special case for
which the universe appeared to be accelerating from a 4d point of
view.

Hence this introduces the following interesting possibility. Here
we may consider the transition our four-dimensional dS space to a
higher-dimensional dS space. However, in this case, we also expect
the reverse process to be possible.  Moreover, we expect that, if
the system can indeed be thought of as a thermal ensemble and the
de Sitter entropies as accurate representations of the number of
states, the solution that dominates the ensemble should be that
with the higher entropy.  This could be either the
higher-dimensional or lower-dimensional de Sitter
space\refs{\Bousso}, depending on the relative magnitudes of the
entropies
\eqn\sten{{\bf S}_{d+4}\sim {M_P^{(d+4)(d+2)/2}\over \Lambda_{d+4}^{(d+2)/2}}}
and
\eqn\sfour{{\bf S}_4\sim {M_4^4\over \Lambda_4}\sim {M_P^{2(d+2)}V_d^2\over \Lambda_4}\ ,}
where $V_d$ is the volume of the compact manifold  at the
``metastable'' minimum. In fact, it is not difficult to construct
examples where the compactified solution has the larger
entropy\refs{\Bousso}. In particular, we know that a minimum
corresponding to the present universe has an entropy ${\bf
S}_4\sim 10^{120}$, and so if the hypothetical higher-dimensional
theory had a cosmological constant that is order unity in Planck
units, our four-dimensional metastable minimum would be expected
to strongly dominate such an ensemble. Hence this scenario could
provide a mechanism for compactification from the
higher-dimensional theory.

Certainly, this is not the entire story. Typically within this
framework, compactified anti-de Sitter solutions will also arise,
\eg\ by adjusting fluxes\refs{\Bousso}. It is not clear what their
role should be in such a scenario\refs{\BanksES,\DineFW,\heretics}. However, this aspect of the dynamics seems a fruitful ground for future research.

\bigskip\bigskip\centerline{{\bf Acknowledgments}}\nobreak

The authors would like to thank R. Bousso,  G. Horowitz, L.
Kofman, A. Linde, M. Lippert, L. Susskind, and M. Taylor for very
valuable conversations.  We also thank Jordan Hovdebo for his
advice in preparing the phase plane figures. The work of SBG was
supported in part by the Department of Energy under Contract
DE-FG02-91ER40618. This work was initiated at the Perimeter
Institute, which SBG would like to thank for its kind hospitality.
Research at the Perimeter Institute is supported in part by funds
from NSERC of Canada. Part of this work was also carried out
during the Superstring Cosmology workshop at the Kavli Institute
for Theoretical Physics, whose support is gratefully acknowledged;
research at the KITP was supported in part by the National Science
Foundation under Grant No. PHY99-07949.

\appendix{A}{Phase plane analysis for $k=\pm1$}

Following the phase plane analysis of \refs{\CopelandET}, we
describe the general solutions of the equations \constrainteq-\phieqn\ for
the cases where the spatial metric is either positively or
negatively curved. Hence in equations \constrainteq\ and \aeqn, we have
$\gamma=2/3$ and substitute for $\rho_\gamma$ as in equation
\rhoevol. The phase plane dynamics explicitly reveals the presence
of fixed point `tracker' solutions, discussed in the main text, as
the generic end-points of various solutions.

\Ifig{\Fig\comboone}{Fixed points and some typical trajectories
for $k=-1$ in the regimes: (i) $\alpha<1$, (ii) $1<\alpha<\sqrt3$,
and (iii) $\alpha>\sqrt{3}$. (The specific plots above were made
for $\alpha=1/\sqrt{3},\ 5/2\sqrt3$, and $2$.)}{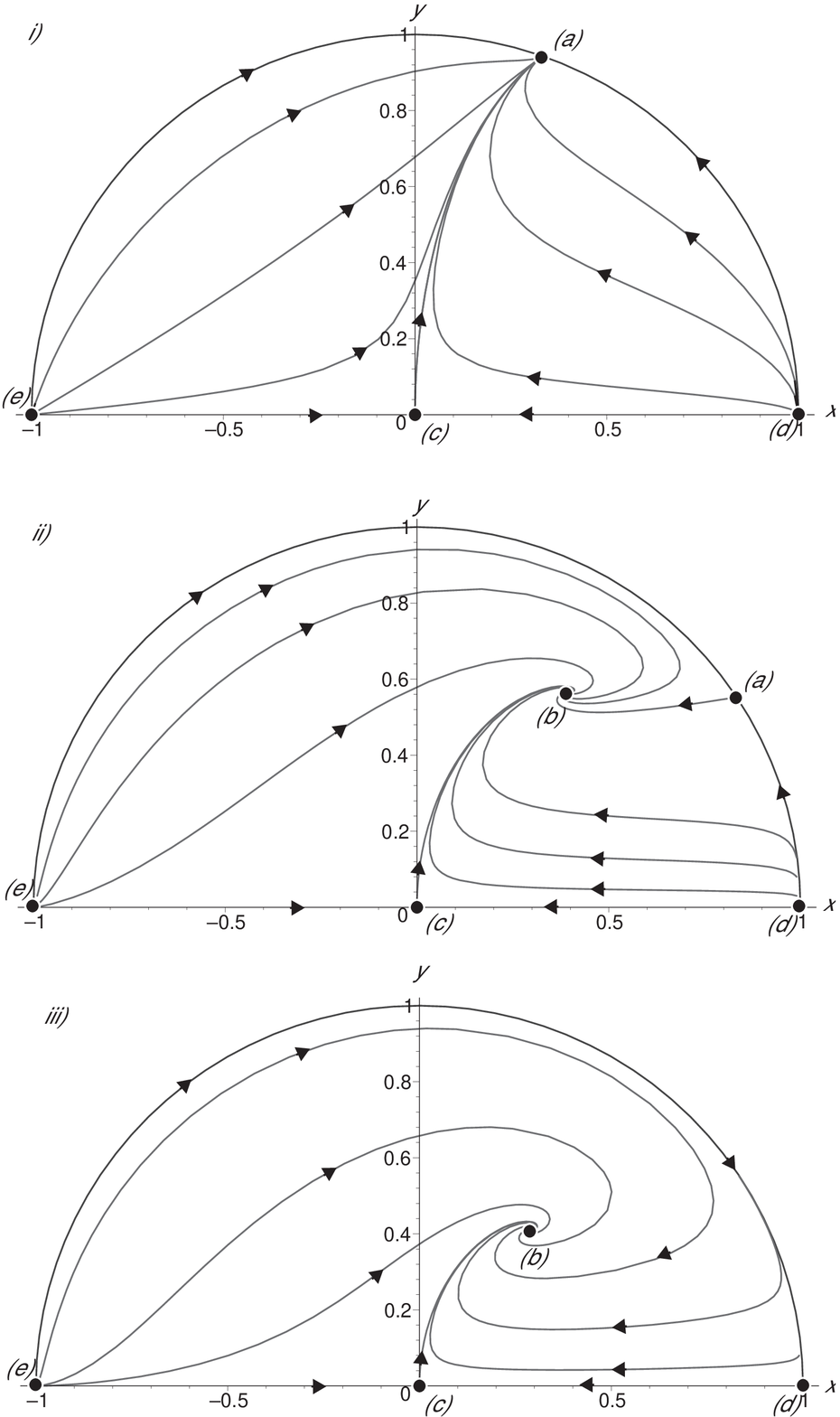}{4.7}

We begin by defining
\eqn\xyzdeff{ x= { {\dot \phi}\over 2{\sqrt 3} \dot{a}/a}
\ ,\qquad
y={{\sqrt V} \over {\sqrt 6} \dot{a}/a}\ ,\qquad
z={1\over {\dot a}}\ .}
With this choice of variables, the Freedman constraint
\constrainteq\ becomes
\eqn\constr{x^2 + y^2 - k\, z^2 = 1\ }
which can be used to eliminate $z$. The second-order equations
\aeqn\ and \phieqn\ can be written in the form:
\eqn\xeqn{ x' = -2x(1-x^2) + y^2 \left(\sqrt{3}\alpha -x\right)\
,}
\eqn\yeqn{y'=y\left(1-y^2\right) -xy\left(\sqrt{3}\alpha-2
x\right)\ ,}
where prime denotes $d/d\log a$ and $\alpha$ is the (positive)
constant characterizing the exponential potential, as in equation
\exppot. Note that these equations are independent of $k$. The
feature which then distinguishes these two cases is the range of
the variables as determined by the constraint \constr, \ie\
$x^2+y^2\le1$ for $k=-1$ and $x^2+y^2\ge1$ for $k=+1$. We are also
generally interested in solutions describing an expanding universe
and so we focus our attention on $y\ge0$.  Since eqns.~\xeqn,
\yeqn\ are symmetric under $y\rightarrow -y$, the $y<0$ dynamics
is a simple extrapolation.

\Ifig{\Fig\combotwo}{Fixed points and some typical trajectories
for $k=+1$ in the regimes: (i) $\alpha<1$, (ii) $1<\alpha<\sqrt3$,
and (iii) $\alpha>\sqrt{3}$. (The specific plots above were made
for $\alpha=1/2,\ 5/2\sqrt3$, and $2$.)}{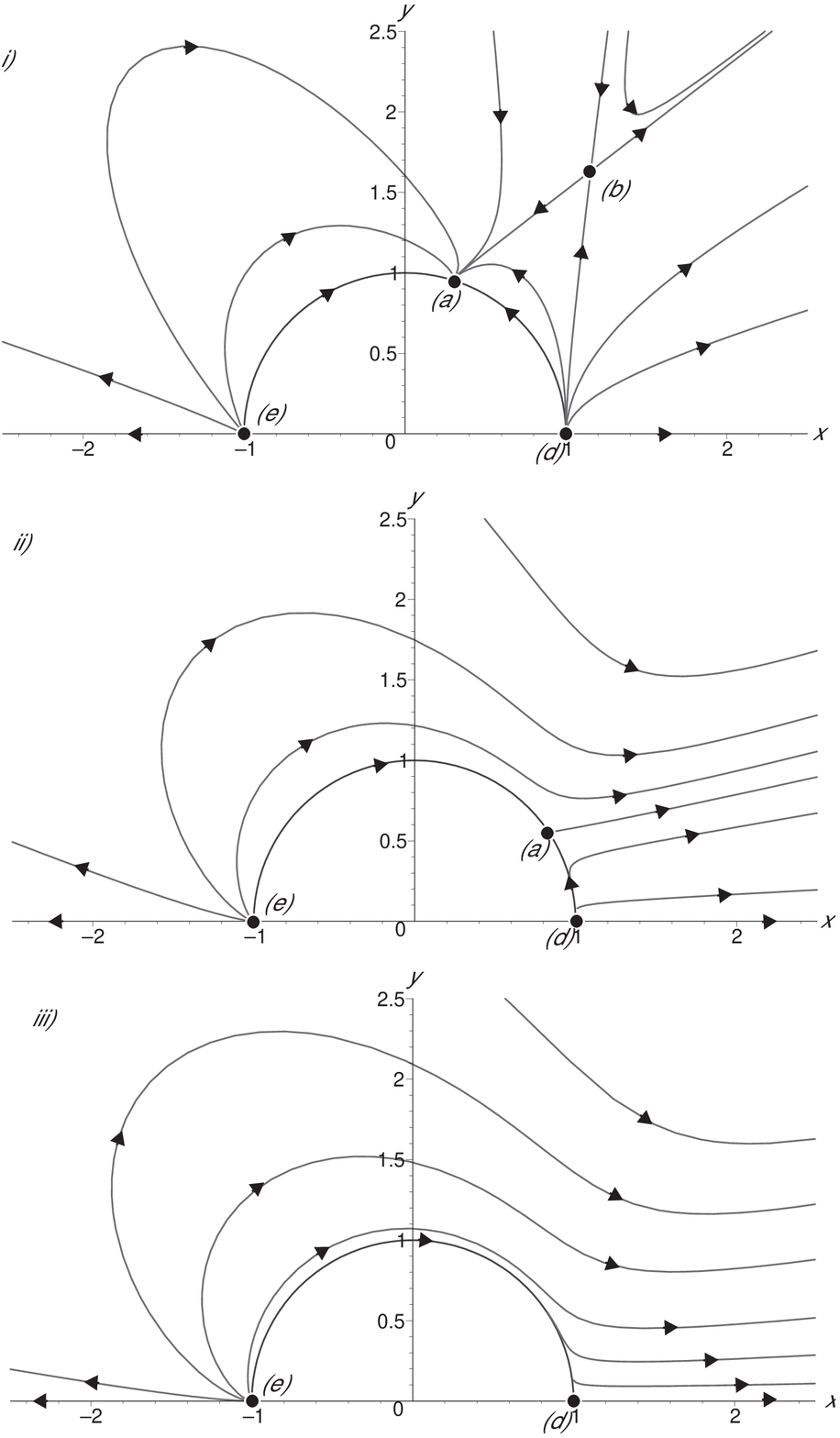}{4.7}

These equations exhibit the following fixed points:

\item{a.} $x=x_a=\alpha/\sqrt{3}$, $y=y_a= \sqrt{1-\alpha^2/3}$,
$z^2=0$,

\item{b.} $x=x_b=1/\sqrt{3}\alpha$, $y=y_b=\sqrt{2}/\sqrt{3}\alpha$,
$z^2=k\left(\alpha^{-2}-1\right)$,

\item{c.} $x=0$, $y=0$, $z^2=-k$,

\item{d.} $x=1$, $y=0$, $z^2=0$,

\item{e.} $x=-1$, $y=0$, $z^2=0$.

\noindent Which of these fixed points is physically realized (and
their stability properties) depends on the values of parameters
$k$ and $\alpha$. In particular, the fixed point (a) only appears
for $\alpha<\sqrt{3}$ in order that $y_a$ is real. While (b) has
$x_b$ and $y_b$ real for all values of $\alpha$, it is only
relevant for $\alpha<1$ with $k=+1$ and $\alpha>1$ with $k=-1$ so
that the fixed point lies in the appropriate domain of the phase
plane, \ie\ $z^2\ge0$. Similarly, (c) is only relevant for $k=-1$.
Finally the fixed points (d) and (e) always appear relevant
independent of the parameters. Figs.~4,5 illustrate the fixed
points and the flows in the phase plane for the various distinct
parameter ranges.   We now comment briefly on these results.

Consider the case $k=-1$, as described by \comboone. First for
$\alpha<1$, (a) is the only stable fixed point and then describes
the end-point behavior of generic solutions. As described in the
text, amongst the string theory potentials, this situation only
seems to be realized with a completely space-filling brane (with
$p=d+3$), in which case this fixed point corresponds to an
asymptotically de Sitter solution in the full ($d$+4)-dimensional
spacetime. For $1<\alpha<\sqrt{3}$, (a) becomes a saddle-point and
then disappears (\ie\ moves off into the complex plane) for
$\alpha>\sqrt{3}$. Hence in the regime $\alpha>1$ which seems to
be the generic case for string theory, (b) is the stable fixed
point.\foot{Qualitatively, these results apply for the case of any
barotropic fluid with a positive energy density\refs{\CopelandET}
--- the latter would be the dominant energy contribution for the
case $k=0$.}

\combotwo\ illustrates the flows for $k=+1$. In this case, (a) is
again a stable fixed point for $\alpha<1$ and the same comments
given above apply here. In fact, (b) only appears in this
parameter range as well, but it is an unstable saddle-point. Hence
for $\alpha>1$, the generic flows run off to infinity in
directions bounded by $y^2/x^2\le2$. This behavior is also generic
for a broad range of initial conditions when $\alpha<1$.
Physically these flows correspond to expanding cosmologies where
the kinetic energy in the scalar field dominates the potential,
and the spatial curvature leads to decelerated expansion.  As a
result $\dot{a}=1/z$ reaches zero at some finite value of $a$ (and
$t$). Hence beyond this point, they begin collapsing. In the phase
plane, this transition is realized by mapping the asymptotic flow
with $(x,y)$ to $(-x,-y)$ and following the flow backwards  with
respect to the arrows indicated in the figures. In the present
context, the solutions of interest in the present context are
those which run off to infinity with $x>0$, \ie\
$\dot{a},\dot{\phi}>0$. In this case, for any value of $\alpha$,
they flow towards a big crunch at the fixed point (e). However, as
commented in the main text, we expect other terms, \eg\ matter and
perturbations, to become important in the final stage of this
crunch.

Note that during the collapse stage above, we have $\dot{a},x<0$
and hence $\dot{\phi}>0$. That is, the internal space continues to
expand for these solutions. From a higher-dimensional perspective,
however, we would still regard this as a collapsing phase, as
given the full metric \tendmet, one can easily show that the
proper volume element is contracting both in the noncompact
four-dimensions and in the full ($d$+4)-dimensional solution.

We have focussed on the end-point of the decompactification
trajectories, but it is interesting to consider the full
trajectories and follow the flows backwards to an initial
fix-point. Generically, for the entire range of parameters, the
flows originate at one of the repulsive fixed points (d) or (e).
Hence these solutions emerge from a singular big bang. As with the
crunch above, the analysis provided here is expected to be
incomplete as many other terms will be important near this initial
singularity. The generic appearance of these singularities was, of
course, observed by \refs{\BanksES,\DineFW,\heretics} and used as
part of their criticism of string theory landscape picture.
However, for tunneling from a metastable dS vacuum, these
singularities don't appear as part of the evolution of the system.
Another possible viewpoint on this may be to consider an
``extended" landscape, in which the 4d radius (or even more
general metric degrees of freedom) is included; in this space the
singular regions may be well separated from the solutions of
interest, {\it e.g.,} by being in different basins of attraction,
though this question deserves further exploration.

We might also comment that, in any range of parameters considered
above, there are many trajectories which begin with negative $x$
and cross over to positive $x$. These would fall into the class of
M-theory solutions, which were recently
discussed\refs{\TownsendRV\Townsend\Emparan\Cornalba\OhtaA\Roy\Wohlfarth\Chen\accel-\OhtaB}
because they exhibit an accelerating phase. As first noted in
\refs{\Emparan}, near $x=0$, the scalar potential momentarily
dominates the energy density and so cosmic acceleration is seen by
four-dimensional observers. However, this is only a momentary
phase since, as noted in the main text, generically the M-theory
potentials are two steep to generate an (continuously)
accelerating solution.

\listrefs
\end